\documentclass[prl,aps,twocolumn]{revtex4}

  \usepackage{amsmath} 
  \usepackage{amssymb} 
  \usepackage{amsthm}
  
\usepackage{graphicx}
  \usepackage[usenames,dvipsnames]{color} 
\usepackage{hyperref}

\begin{document}

\title{Comment on ``Nonlinear fluctuations and dissipation in matter revealed by quantum light''}
\author{M.~Kira,$^{1,\star}$, S.W.~Koch$^1$, and S.T.~Cundiff$^2$}

\affiliation{$^{1}$Department of Physics, Philipps University Marburg, Renthof 5, D-35032 Marburg, Germany}

\affiliation{ $^{2}$Physics Department, University of Michigan, Ann Arbor, Michigan 48109, USA}

\normalsize{$^\ast$To whom correspondence should be addressed; email:  kira@staff.uni-marburg.de}

\date{\today}

\begin{abstract}
In a recent paper [Phys.~Rev.~A {\bf 91}, 053844 (2015)], 
Mukamel and Dorfman compare spectroscopies performed with classical vs.~quantum light, 
and conclude that \textit{nonlinear} quantum-spectroscopy signals cannot be obtained from averaging 
their classical-spectroscopy counterparts over the Glauber--Sudarshan quasiprobability
distribution of the quantum field.
In this Comment, we show that this interpretation is correct only if one assumes that a classical signal is given by 
a classical approximation for the field. While such an assumption can be useful for comparing theoretical results, it is never realized in laser spectroscopy experiments that typically use coherent states. Thus, instead of using classical signals, the connection between coherent states and quantum states of light must be considered. We rigorously show that quantum spectroscopy can always be projected from the experimentally realized coherent-state spectroscopy regardless how nonlinear the system response is.
\end{abstract}

\maketitle

\newcommand{\op}[1]{\hat{#1}}
\newcommand{\ave}[1]{\langle #1 \rangle}
\newcommand{\comm}[2]{\left[ #1,\,#2 \right]_{-}}
\newcommand{\trace}[1]{{\rm Tr}\left[ #1 \right]}
\newcommand{\eps}{\varepsilon}

\newcommand{\be}{\begin{equation}}
\newcommand{\ee}{\end{equation}}
\newcommand{\bea}{\begin{eqnarray}}
\newcommand{\eea}{\end{eqnarray}}
\newcommand{\ev}[1]{\langle#1\rangle}
\newcommand{\ddt}{\frac{\partial}{\partial t}}
\newcommand{\ihddt}{i\hbar\frac{\partial}{\partial t}}
\newcommand{\mcal}[1]{{\mathcal{#1}}}
\newcommand{\mrm}[1]{{\mathrm{#1}}}
\newcommand{\drm}{{\mathrm{d}}}
\newcommand{\e}{\mathrm{e}}
\renewcommand{\matrix}[1]{\mathbf{#1}}
\renewcommand{\vec}[1]{\mathbf{#1}}
\newcommand{\ci}[1]{\mathbf{#1}}
\newcommand{\stumm}{\bullet}
\newcommand{\ssum}[1]{\left[\left[#1\right]\right]}
\newcommand{\csum}[2]{\left|\left|#2\right|\right|_{#1}}
\newcommand{\widebar}[1]{\overline{#1}}
\newcommand{\bra}[1]{{\langle #1 |}}
\newcommand{\ket}[1]{{| #1 \rangle}}

Finding and understanding the border between the classical and quantum regimes is essential for establishing the foundations of modern optics. For example, the seminal contributions by Glauber\cite{Glauber:1963} and Sudarshan\cite{Sudarshan:1963} identified coherent states, $\ket{\beta}$, as the most classical form of quantized light. Once the phase of a single-mode light field, such as is emitted by a perfect laser, is defined against a reference (otherwise light's phase remains random, as pointed out by M{\o}lmer\cite{Molmer:1997}), it is quantum statistically described by one coherent state. Normally-ordered expectation values for $\ket{\beta}$ are obtained by a classical factorization\cite{Book:2011,Walls:1994} where the photon operators $\op{B}$ and $\op{B}^\dagger$ are replaced by complex amplitudes $\beta$ and $\beta^\star$, respectively, as if the field had a definite classical phase and amplitude. However, coherent states have quantum fluctuations that invalidate the classical factorization for 
expectation values that are not normally ordered.

One can use coherent states to express the density matrix of a single-mode light via\cite{Glauber:1963,Sudarshan:1963}
\begin{eqnarray}
  \op{\rho} = \int d^2 \beta\, P(\beta) \,\ket{\beta} \bra{\beta}
\label{eq:Pdecomposition}\,,
\end{eqnarray}
where $P(\beta)$ is the Glauber--Sudarshan function. This relation is the starting point of the analysis by Mukamel and Dorfman\cite{Mukamel:2015}. As pointed out by Glauber and Sudarshan, Eq.~\eqref{eq:Pdecomposition} appears to be a probabilistic phase-space average over coherent states, i.e.~$\ket{\beta}\bra{\beta}$.  For a light described by $\ket{\beta_0}$, $P(\beta)$ becomes $\delta(\beta-\beta_0)$. States of light with a positive-definite $P(\beta)$ must necessarily have equal or greater fluctuations compared to coherent states. Since their fluctuations result from a probabilistic average of the coherent states, these states of light are often referred to as ``semiclassical``. There are also many states of light, such as Fock and Schr\"{o}dinger-cat states, whose $P(\beta)$ is partially negative valued or even nonanalytic. Such sources are often referred to  as ``true quantum light``. 

At the moment, there exist very few experimental realizations of true quantum light, which are technically challenging and usually limited to very low intensities. Therefore, it is interesting to determine whether spectroscopy performed with semiclassical vs.~true quantum sources are connected. In this context, experiments performed with coherent states have been labeled as ``classical spectroscopy'', whereas those performed with true quantum light have been labeled ``quantum (optical) spectroscopy''\cite{Kira:2006b,Kira:2011}. However, it is important to bear in mind that since ``classical spectroscopy'' refers to measurements made with coherent states, there are still quantum aspects to the measurement, namely the intrinsic quantum fluctuations, and the results of these measurements are not necessarily the same as for using the approximation of purely classical light, which is actually not experimentally possible. We will try to emphasize this point by avoiding the term ``classical spectroscopy'' and 
explicitly stating when we are referring measurements made with coherent states.

In his 1963 paper, Sudarshan essentially writes a projection between outcomes for coherent-states and quantum light \cite{Sudarshan:1963}. This projection can be used to establish a connection between quantum spectroscopy and coherent-state spectroscopy. Using the notation from Refs.~\cite{Kira:2011,Mootz:2014,Hunter:2014,Berger:2014}, the system response to quantum light, $R_{\rm QM}$, in terms of the responses to coherent-states, $R_\ket{\beta}$, is 
\begin{eqnarray}
  R_{\rm QM} = \int d^2 \beta\, P(\beta) \, R_\ket{\beta}
\label{eq:Pprojection}\,,
\end{eqnarray}
where the quantum-light statistics are defined by $P(\beta)$. Sudershan defines $R_\ket{\beta}$ via 
normally ordered operators, and we show later that, in general, the outcome of a spectroscopy measurement can eventually be expressed via $R_\ket{\beta}$. In the case where a spectroscopic technique uses multiple single-mode sources or a multi-mode source, the connection between coherent-state and quantum-optical measurements, \eqref{eq:Pprojection}, can be generalized by replacing $\beta$ by $\beta_1, \beta_2$ etc., where $\beta_j$ defines the quantum statistics of each orthogonal mode applied in the technique. We will next study whether the claims made by Mukamel and Dorfman\cite{Mukamel:2015} that
Eq.~\eqref{eq:Pprojection} ``only holds for the linear response, and does not apply to the
nonlinear response'' and ``The quantum response carries additional valuable
information about response and spontaneous fluctuations of matter that may not be retrieved
from the classical response by simple data processing'' are justified.

In their paper, Mukamel--Dorfman\cite{Mukamel:2015} consider two examples where $R_\ket{\beta}$ is first computed and then it is analyzed how the projection \eqref{eq:Pprojection} connects different levels of calculations. In the first example, they discuss the third-order nonlinear spectrum of a system which they denote as $S(\omega)$. 
They apply a classical factorization, i.e.~they replace the photon operators by complex-valued amplitudes, to obtain $S(\omega)_{\rm cl}$ which they treat as the system response to a classical field in Eq.~(14) of Ref.~\cite{Mukamel:2015}. 
They compare this result with a calculation that does not apply the classical factorization, giving the quantum response $S(\omega)_{\rm QM}$. They correctly find that $S_{\rm QM}$ cannot be reproduced by the projection \eqref{eq:Pprojection} when $R_\ket{\beta}$ is replaced by $S_{\rm cl}$. Furthermore, they argue that this violation results because ``a classical field does not change in the course of field--matter interactions''.

To analyze this apparent violation of the quantum-spectroscopy concept\cite{Kira:2006b,Kira:2011}, it is helpful to consider whether a laser-spectroscopy experiment performed with a coherent state, actually yields $S_{\rm cl}$ ($R_{\rm cl}$ in the notation we are using) 
resulting from a classical factorization. 
In this context, it is a very instructive example is to consider the laser excitation of a two-level system inside a cavity. Using the Jaynes--Cummings model\cite{Jaynes:1963} to compute the system response $R_\ket{\beta}$ to a coherent state $\ket{\beta}$ {\it without} the classical factorization shows that the occupation of the excited state first increases but then collapses\cite{Cummings:1965} even without damping. It was also found in 1960s that the occupation collapse is followed by a series of revivals and further collapses \cite{Eberly:1980}. However,
as soon as a classical factorization is applied, the theory reduces to the optical Bloch equations\cite{Allen:1987,Book:2011} that produce the occupation $R_{\rm cl}$ corresponding to undamped Rabi oscillations when dephasing is excluded. Consequently, the response, $R_\ket{\beta}$, to a coherent state is drastically different from $R_{\rm cl}$ obtained with a classical factorization. Obviously, the usefulness of $R_\ket{\beta}$ vs.~$R_{\rm cl}$ can be judged based on
which of them predicts the outcome of an actual laser-spectroscopy measurement. In 1987, Rempe {\it et al.}\cite{Rempe:1987} used a maser to measure a clear sequence of collapses and revivals, which demonstrates that a maser/laser spectroscopy measures $R_\ket{\beta}$, not $R_{\rm cl}$.

This conclusion is not surprising because a measurement cannot switch off the quantum aspects of a coherent state --- at its best, a laser is in maximally classical coherent state $\ket{\beta}$ whose quantum fluctuations still influence the measured response $R_\ket{\beta}$. As originally formulated by Sudarshan\cite{Sudarshan:1963}, the projection relation \eqref{eq:Pprojection} connects $R_\ket{\beta}$, not $R_{\rm cl}$, with a quantum-optical response, $R_{\rm QM}$, to a true quantum-light source. In our quantum-spectroscopy investigations\cite{Kira:2011,Mootz:2014,Hunter:2014,Berger:2014}, we carefully apply this fundamental connection, and correctly use $R_\ket{\beta}$ as the system response to a coherent state, obtained without any classical approximations to the field and response.

By using classically approximated $R_{\rm cl}$ instead of the full $R_\ket{\beta}$, Mukamel and Dorfman\cite{Mukamel:2015} conclude that the projection \eqref{eq:Pprojection} ``breaks down in the nonlinear regime''. Therefore, it is instructive to check whether this conclusion holds for spectroscopy using coherent states. Before the light excites matter, it can be assumed not to be entangled with the matter such that the initial state of the total system can be expressed as $\hat{\rho}_{\rm tot}(0) = \hat{\rho}_{\rm mat} \otimes \hat{\rho}$ where $\hat{\rho}_{\rm mat}$ is the initial state of matter while $\hat{\rho}$ defines the light state. For laser spectroscopy,  $\hat{\rho}$ becomes $\ket{\beta}\bra{\beta}$ of a coherent state, which yields $\hat{\rho}_\ket{\beta}(0) \equiv \hat{\rho}_{\rm mat} \otimes \ket{\beta}\bra{\beta}$ as the initial state of the combined matter-light system. When this result and Eq.~\eqref{eq:Pdecomposition} are merged, the initial state becomes
\begin{eqnarray}
  \hat{\rho}_{\rm tot}(0) = \int d^2 \beta\, P(\beta) \, \hat{\rho}_\ket{\beta}(0)
\label{eq:rho_ini}\,.
\end{eqnarray}
For a closed system consisting of only the matter and the exciting light, the full evolution follows from the system Hamiltonian $\hat{H}$ including the quantized light--matter coupling as well as possible many-body interactions within the matter. In the Schr\"{o}dinger picture, the exact time evolution formally follows from 
$\hat{\rho}_{\rm tot}(t) = e^{-i \hat{H} t/\hbar}\, \hat{\rho}_{\rm tot}(0) \, e^{+i \hat{H} t/\hbar}$,
regardless how nonlinear the interactions are. By inserting Eq.~\eqref{eq:rho_ini} into this relation, we find
\begin{eqnarray}
  \hat{\rho}_{\rm tot}(t) = \int d^2 \beta\, P(\beta) \, \hat{\rho}_\ket{\beta}(t)
\label{eq:rho_t}
\end{eqnarray}
because $P(\alpha)$ is not an operator. In this decomposition, $\hat{\rho}_\ket{\beta}(t)$ corresponds to the {\it exact} evolution following an excitation by a coherent-state $\ket{\beta}$. It is clear that the part of $\hat{\rho}_\ket{\beta}(t)$ describing the light does not necessarily remains a pure $\ket{\beta}$ when interactions are present, which invalidates the classical factorization. Therefore, \eqref{eq:rho_t} is no longer valid when the classical factorization is applied to approximate the exact $\hat{\rho}_\ket{\beta}(t)$.

The general system response $R_{\rm QM}$ corresponds to an expectation value of an observable $\op{R}$,
i.e.~$R_{\rm QM} \equiv \ave{\op{R}} = {\rm Tr} \left[ \op{R} \, \hat{\rho}_{\rm tot}(t) \right]$, or a Fourier transform over $t$ yielding spectral information.
By using the connection \eqref{eq:rho_t}, we find
\begin{align}
  R_{\rm QM} &= {\rm Tr} \left[ \op{R} \, \int d^2 \beta\, P(\beta) \, \hat{\rho}_\ket{\beta}(t) \right]
\nonumber\\
  &= \int d^2 \beta\, P(\beta) \,{\rm Tr} \left[ \op{R} \,  \hat{\rho}_\ket{\beta}(t) \right]
\label{eq:R_t}\,,
\end{align}
which follows by exchanging the trace an integration. We notice now that ${\rm Tr}\left[ \op{R} \,  \hat{\rho}_\ket{\beta}(t) \right]$ is the exact system response $R_\ket{\beta}$ to a coherent state $\ket{\beta}$.
Therefore, Eq.~\eqref{eq:R_t} reduces to an exact projection \eqref{eq:Pprojection} even when the spectroscopic response is nonlinear. The opposite conclusion by Mukamel and Dorfman\cite{Mukamel:2015} stems from the classical factorization which, however, is not applicable to describe real laser spectroscopy. We have shown in Ref.~\cite{Kira:2011} that the projection \eqref{eq:Pprojection} is also valid for open systems. Basically, the projection is a direct consequence of
the linearity of quantum mechanics, irregardless of the nonlinearity of $R_\ket{\beta}$ induced by many-body and quantum-optical interactions.

In their second example, Mukamel and Dorfman\cite{Mukamel:2015} investigate nonlinear fluctuation--dissipation relations with superoperators. In their Eq.~(19), the system response is ${\rm Tr}\left[O_1 \rho(t)\right]$ where $O_1$ correspond to our $\op{R}$ and $\rho(t)$ to $\hat{\rho}_{\rm tot}(t)$ converted to the interaction picture. They apply a classical factorization when computing ${\rm Tr}\left[O_1 \rho(t)\right]$ and conclude that ``The deep reason why the CRF and QRF are not simply
related is in the lack of a fluctuation-dissipation relation in the nonlinear regime'' where CRF (QRF) stands for the classical (quantum) response function.
However, a laser spectroscopy experiment does not measure ${\rm Tr}\left[O_1 \rho(t)\right]$ obtained with a classical factorization for the field, but rather the full ${\rm Tr}\left[ \op{R} \,  \hat{\rho}_\ket{\beta}(t) \right]$ response to a coherent state. As we show above, the projection \eqref{eq:Pprojection} connects laser spectroscopy performed with coherent states and quantum spectroscopy.

Since experiments realizing laser spectroscopy with coherent states are common place whereas 
the technology for generating arbitrary states of quantum light is still primitive, the
projection \eqref{eq:Pprojection} yields an interesting method for deducing quantum responses from a large set of laser-spectroscopy measurements using coherent states. However, many interesting quantum sources have an extremely complicated $P(\beta)$ function consisting, e.g., of high-order derivatives of a delta function. 
Since $R_\ket{\beta}$ cannot be measured experimentally throughout the entire phase-space and any measurement inevitably contains some additional noise, it is clear that integrating  \eqref{eq:Pprojection}  with experimental data and a rapidly changing (or nonanalytic) quantum $P(\beta)$ is challenging.

We have devised a numerical algorithm\cite{Kira:2011}, based on a cluster-expansion transformation (CET)\cite{Kira:2008},
to robustly 
integrate $\int d^2 \beta\, P(\beta) \, R_\ket{\beta}$ from a finite number of $\ket{\beta}$ measurements.
We have thoroughly tested our CET projection in Ref.~\cite{Kira:2011} with the Jaynes-Cummings model,
which is one of the few nontrivial examples where $R_\ket{\beta}$ can be computed exactly. More specifically,
we computed  $R_\ket{\beta}$ for 240 $\beta$ values for $0 < |\beta| <9$ resulting to the well-known collapse-revival sequences explained above. We then applied our CET projection to determine $R_{\rm QM}$ to Fock-states as quantum sources and fully recovered the related quantum-Rabi oscillations, which also have been measured\cite{Brune:1996}. In the same context, we have also shown\cite{Kira:2011} that the CET projection method works even when random noise is added to the data demonstrating that
our method is robust against experimental noise.
Hence, it can be directly applied to project coherent-state laser spectroscopy measurements to true quantum-optical spectroscopy.

We have also shown in Ref.~\cite{Kira:2011} that the quantum and classical spectroscopies produce different results only when $R_\ket{\beta}$ exhibits nonlinearities. Therefore, 
the difference of $R_{\rm QM}$ and $R_\ket{\beta}$ 
is a particularly sensitive method for detecting many-body and quantum-optical effects that produce nonlinearities. We have applied the CET projection to detect complex many-body quasiparticles such as biexcitons\cite{Mootz:2014} and dropletons\cite{Hunter:2014} in a semiconductor quantum well as well as a photon-density correlation\cite{Berger:2014} in the semiconductor quantum-dot emission, and we expect our method to be an extremely useful extension of laser spectroscopy into quantum spectroscopy.

In conclusion, Mukamel and Dorfman\cite{Mukamel:2015} analyze how different levels of theory compare with one another, and we fully agree that a computation performed with a classical factorization does not contain the quantum-statistical information needed to project quantum-optical responses. However, a laser-spectroscopic measurement cannot be described by the classical factorization since it cannot switch off the quantum fluctuations of light, such that it always contains the quantum-statistical information required to validate the projection \eqref{eq:Pprojection}. We have shown that laser spectroscopy performed using coherent states connects with measurements performed with true quantum sources via the projection \eqref{eq:Pprojection}, which is extremely useful because one can routinely realize coherent-state spectroscopy while quantum-
source development is challenging.
Even though quantum spectra can be projected from a massive set of coherent-state spectroscopy data, it is not any less exciting. One may view that a set of laser excitations disperse the full many-body and quantum-optical information between all possibilities, whereas quantum spectroscopy condenses this information to an isolated quantum response. Analogous ``parallelization'' of outcomes makes quantum computing interesting compared to its classical counterparts.

We thank Shaul Mukamel for discussions and providing us a preprint of Ref.~\cite{Mukamel:2015}. The work in Marburg is supported by the Deutsche Forschungsgemeinschaft (grant KI 917/2-2).


\end{document}